\documentclass[conference]{IEEEtran}
\IEEEoverridecommandlockouts
\usepackage{cite}
\usepackage{amsmath,amssymb,amsfonts}
\usepackage{algorithmic}
\usepackage{graphicx}

\usepackage{threeparttable}
\usepackage{tabularx}
\usepackage{booktabs}
\usepackage{balance}
\usepackage{textcomp}
\usepackage{xcolor}
\def\BibTeX{{\rm B\kern-.05em{\sc i\kern-.025em b}\kern-.08em
    T\kern-.1667em\lower.7ex\hbox{E}\kern-.125emX}}
\begin{document}

\title{EscFOA: Enhancing Spatial Learning for Visually Impaired Learners via Generative Spatial Audio in 360-Degree Educational Environments}

\author{
\IEEEauthorblockN{
Ziyu Luo$^{*}$,
Xiaowei Dai$^{*}$,
Siying Zhu$^{\dagger}$ (Correspondence),
Xiaoming Chen$^{*}$ (Correspondence)
\thanks{This work was supported by National Natural Science Foundation of China (NSFC) under Grant 62577004.}
}
\IEEEauthorblockA{
$^{*}$School of Computer and Artificial Intelligence, Beijing Technology and Business University, Beijing, China\\
Emails: \{2431062101, 2330702014\}@st.btbu.edu.cn, xiaoming.chen@btbu.edu.cn\\
$^{\dagger}$College of Fundamental Studies, Beijing Vocational College of Labour and Social Security, Beijing, China\\
Email: 2018010308@bvclss.edu.cn
}
}

\maketitle

\begin{abstract}
Immersive 360-degree educational environments often lack accessible spatial structure, limiting visually impaired learners' ability to orient, explore, and construct mental representations. This paper proposes EscFOA, a geometry-aware spatial audio generation framework designed as an \emph{acoustic scaffolding} to support spatial cognition. By integrating 3D Gaussian Splatting (3DGS) with conditional diffusion models, EscFOA reconstructs scene geometry from 360-degree videos to synthesize high-fidelity spatial audio consistent with the environmental structure. Explicitly targeting learning outcomes like independent spatial orientation and reduced cognitive load, EscFOA significantly outperforms conventional monaural and stereo audio in supporting spatial learning behaviors among blindfolded sighted participants (simulating visually impaired learners). These findings demonstrate that geometry-consistent generative audio can effectively enable inclusive access to complex spatial learning materials.
\end{abstract}


\section{Introduction}
\begin{figure}[t]
\centering
\includegraphics[width=0.7\columnwidth]{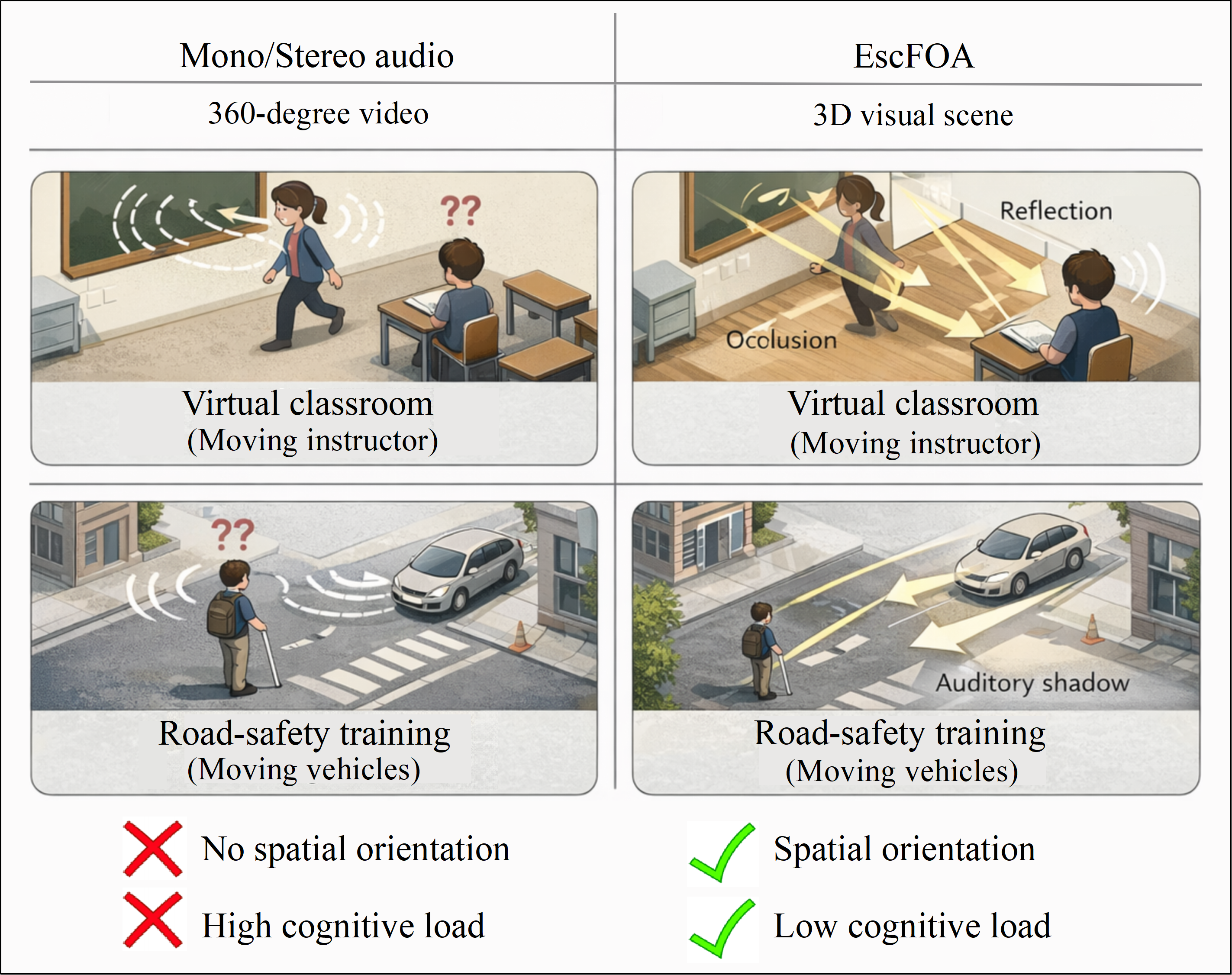}
\caption{\textbf{Overview of EscFOA for enhanced spatial learning.} Left: conventional mono/stereo audio provides limited and unstable spatial landmarks for visually impaired learners in 360-degree educational environments. Right: EscFOA generates geometry-aware spatial audio to support spatial orientation and reduce cognitive load in virtual classrooms, for example, road-safety scenarios.}
\label{fig:teaser}
\end{figure}

According to the World Health Organization (WHO), approximately 2.2 billion people worldwide are affected by varying degrees of vision impairment, among whom 217 million are classified as having moderate to severe vision impairment~\cite{who2023blindness,chen2025vf}. In digital educational environments, spatial understanding is essential for visually impaired learners to interpret, organize, and internalize instructional content. However, in immersive educational environments, they often have limited access to spatial information despite rich audiovisual content. Although speech remains accessible, critical cues such as the instructor’s movement trajectories are often missing. Lahav et al.~\cite{lahav2012virtual} identified such cues as necessary for building accurate internal spatial representations of the environment. In high-stakes tasks such as urban traffic safety training, learners must also interpret ``auditory shadows'' and reflections from facades to identify vehicles approaching from occluded corners~\cite{xu2023wearable}. As illustrated in Fig.~\ref{fig:teaser}, EscFOA aims to provide geometry-consistent auditory landmarks for tracking moving sound sources in immersive educational environments. Such support should extend beyond source localization to include cues such as occlusion, reflection, and reverberation, helping learners infer room layout and spatial relationships during exploration. This challenge is therefore both perceptual and pedagogical, because spatial cognition underpins contextual reasoning, memory organization, and self-directed exploration. Without coherent spatial structure, visually impaired learners are more likely to rely on passive listening, which increases cognitive load and weakens internal spatial representations.

First-Order Ambisonics (FOA) ~\cite{zotter2019ambisonics} is a widely used spatial audio representation that encodes a full-sphere sound field with four channels $(W, X, Y, Z)$, enabling head-rotation-consistent rendering and robust directional perception in 360-degree playback. Yet, many real-world 360-degree videos are still distributed with only monaural or stereo audio because of capture and production constraints, which greatly limits spatial cues such as stable direction, occlusion-induced attenuation, and reflection/reverberation patterns. This limitation is especially problematic for visually impaired learners, who rely more heavily on auditory scene structure to maintain orientation, infer layout, and explore independently when visual spatial structure is inaccessible. Traditional stereo audio often cannot provide sufficiently consistent spatial landmarks to support these learning-relevant behaviors~\cite{picinali2014exploration}.

To address these limitations, we propose EscFOA (Enhancing Spatial Cognition via First-Order Ambisonics), a learning-oriented, geometry-aware generative spatial audio framework that generates FOA from 360-degree educational videos, as shown in Fig.~\ref{fig:teaser}. Inspired by the foundational ``scaffolding'' theory of Wood et al.~\cite{wood1976role}, EscFOA transforms 360-degree visual scene structure into \emph{acoustic scaffolding} for visually impaired learners. Specifically, EscFOA builds on DynFOA~\cite{luo2026dynfoa} by adapting its geometry-aware FOA generation pipeline to immersive educational environments. It uses 3D Gaussian Splatting (3DGS)~\cite{kerbl20233d} to recover scene geometry and coarse material cues, and conditions a diffusion-based generator~\cite{Liu2023AudioLDM} to synthesize FOA whose occlusion, reflection, and reverberation cues are aligned with the surrounding 3D structure. Rather than treating spatial audio as a purely perceptual enhancement, EscFOA targets learning-relevant outcomes, including spatial orientation and independent exploration, by providing stable auditory landmarks and supporting active acoustic probing. Our evaluation compares EscFOA with monaural and stereo baselines in 360-degree educational settings and shows consistent improvements in spatial learning performance.

\section{Related Work}

\subsection{Situated Cognition in Inclusive Education}

Situated Cognition Theory (SCT)~\cite{wilson2000situated} emphasizes that learning is most effective when knowledge is grounded in meaningful physical and social contexts. For visually impaired learners, however, such grounding is often weakened in immersive digital environments: insufficient or poorly structured environmental signals can force passive listening and hinder the construction of accurate spatial mental models~\cite{lahav2012virtual}. Recent work further shows that auditory regularity and environmental coherence are critical for spatial comfort and wayfinding~\cite{zou2023spatial}, yet many immersive educational systems still lack a consistent spatial framework to support situated learning through accessible spatial representations.

\subsection{Acoustic Scaffolding for Visually Impaired Learners}

Wood et al.~\cite{wood1976role} introduced scaffolding as temporary support that helps learners bridge the gap between current ability and learning goals. In assistive settings, however, many existing travel aids for visually impaired learners still focus on obstacle detection or point-based alerts, offering limited support for higher-level spatial understanding~\cite{xu2023wearable}. We therefore frame geometry-consistent spatial audio as \emph{acoustic scaffolding}, where cues such as occlusion, reflection, and reverberation serve as stable landmarks for active reasoning, independent exploration, and gradual internalization of spatial relationships.

\subsection{Generative Artificial Intelligence for Educational Equity}

Generative Artificial Intelligence (AI) offers new opportunities for educational accessibility by automating adaptive media creation. Latent diffusion models enable high-fidelity controllable audio synthesis~\cite{Liu2023AudioLDM}, while 3DGS provides an efficient representation for complex 3D scenes~\cite{kerbl20233d}. However, existing approaches mostly emphasize content generation rather than learning-oriented representation. EscFOA builds on these advances to transform scene geometry into auditory scaffolding for inclusive and cognitively grounded learning.

\section{Methodology}

\begin{figure}[htbp]
\centering
\includegraphics[width=\linewidth]{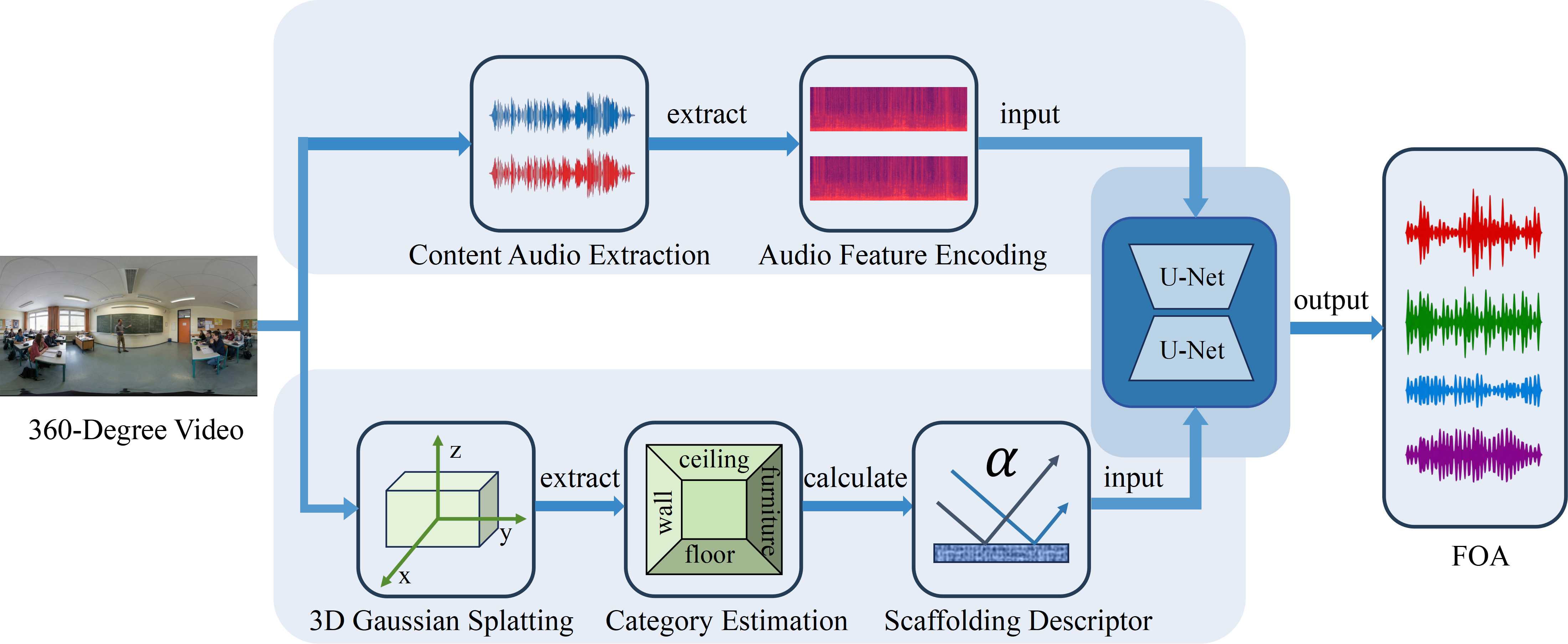}  % 宽度改为单栏宽度
\caption{\textbf{EscFOA: a geometry-aware generative spatial audio framework for learning-oriented ``\emph{acoustic scaffolding}''.} 
EscFOA extracts information from 360-degree videos, computes a scaffolding descriptor, and generates geometry-conditioned FOA. The cues support spatial orientation and independent exploration for visually impaired learners.}
\label{fig:fig1_framework}
\end{figure}

EscFOA is an \emph{acoustic scaffolding} framework for spatial learning rather than a full acoustic simulator. In immersive 360-degree lessons, visually impaired learners often lack stable access to spatial structure, which impairs orientation and independent exploration. To address this issue, EscFOA converts visible scene structure into consistent and learnable auditory cues. As shown in Fig.~\ref{fig:fig1_framework}, the framework first builds a coarse layout with key entities, then derives a compact scaffolding descriptor, and finally generates FOA with spatial cues that remain consistent under head movement. The technical pipeline is derived from DynFOA~\cite{luo2026dynfoa}, which is outlined below.

\subsection{Pipeline}
Starting from a 360-degree educational video, EscFOA uses 3DGS to recover a stable 3D representation of the environment~\cite{kerbl20233d}. Instead of pursuing full physical accuracy, it extracts only learning-relevant elements, such as major walls, ceiling planes, large furniture, and dominant moving sound sources (e.g., an instructor or approaching vehicles). Coarse surface categories, including wall, ceiling, floor, and furniture/obstacle, are assigned to provide stable cues for occlusion and reflection. This design follows the scaffolding perspective: the goal is not exhaustive acoustic simulation, but interpretable support for spatial orientation and engagement.

\subsection{Geometry-aware FOA Generation}
From the reconstructed layout, EscFOA computes a compact scaffolding descriptor composed of learner-to-instructor distance $d_t$, visibility or occlusion $v_t$, and a coarse histogram of nearby surface categories $\mathbf {h}_t$. These variables provide simple but learnable geometry-linked landmarks: $v_t$ controls occlusion-related attenuation, $\mathbf {h}_t$ approximates nearby boundaries that shape early reflections, and $d_t$ stabilizes the direct-to-reverberant relationship during exploration.

EscFOA also supports active acoustic probing, allowing learners to infer wall or obstacle proximity through reflected cues~\cite{picinali2014exploration}. A U-Net-based conditional generative audio model as in DynFOA~\cite{luo2026dynfoa} then synthesizes FOA from content audio, visual context, and this descriptor. During playback, the generated FOA is dynamically rotated based on head tracking and rendered binaurally through the Oculus VR headset's integrated audio system. This provides a practical way to adapt 360-degree educational videos without bespoke acoustic authoring~\cite{Liu2023AudioLDM}.

\section{Experiment}
To evaluate EscFOA in immersive 360-degree educational environments, we conducted a controlled user study focusing on spatial orientation and perceived cognitive load. Following established auditory VR protocols~\cite{picinali2014exploration}, 32 sighted participants (20 male, 12 female; mean age = 25) were blindfolded (simulating visually impaired learners). Participants completed navigation and spatial reasoning tasks in classroom and street scenes from the Sphere360 dataset~\cite{liu2025omniaudio} under three auditory conditions: monaural (Mono), stereo, and EscFOA. During exploration, they could trigger brief acoustic probes to inspect reflected cues~\cite{picinali2014exploration, zou2023spatial}. We report navigation behavior and 5-point Mean Opinion Scores (MOS) for perceived effort and learning.

\begin{figure}[h]
\centering
\includegraphics[width=0.33\textwidth]{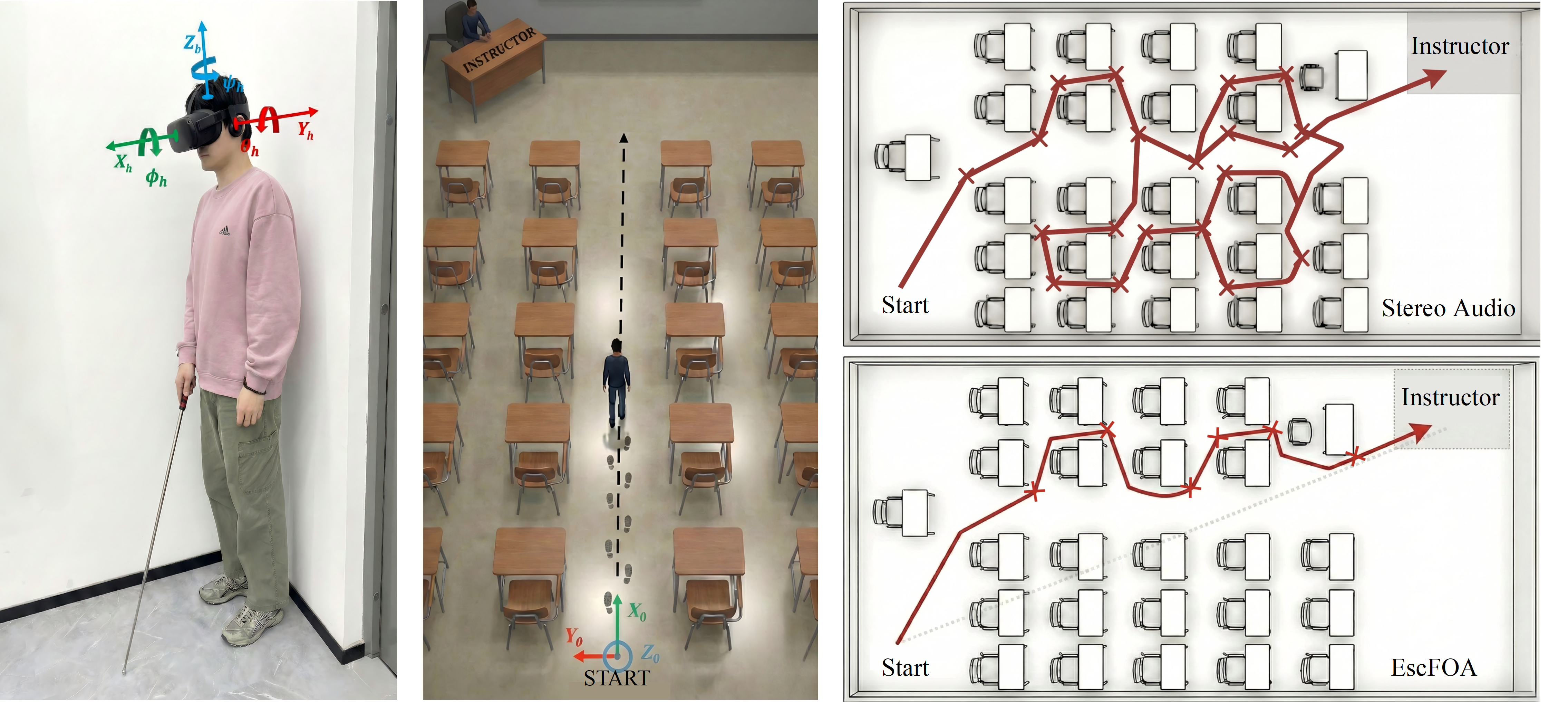}
\caption{\textbf{Navigation experiment for spatial orientation and independent exploration.} Left: a blindfolded participant. Middle: the virtual environment. Right: navigation trajectories under stereo and EscFOA; crosses indicate collisions.}

\label{fig:experiment}
\end{figure}

As illustrated in Fig.~\ref{fig:experiment}, trajectories under EscFOA are smoother and involve fewer collisions than stereo, suggesting more stable orientation and more confident exploration. These qualitative results indicate that geometry-consistent auditory cues can better support spatial navigation under visually deprived conditions.

\begin{table}[htbp]
  \centering
  \caption{MOS RESULTS FOR SUBJECTIVE EVALUATION OF LEARNING EXPERIENCE.}
  \label{tab:mos_results}
  \resizebox{0.90\columnwidth}{!}{%
  \begin{tabular}{lccc}
    \toprule
    Evaluation Metric & Mono & Stereo & \textbf{EscFOA (Ours)} \\
    \midrule
    Perceived Ease             & $2.74 \pm 0.65$ & $3.38 \pm 0.52$ & $\mathbf{4.18 \pm 0.41}$ \\
    Listening Comfort          & $3.15 \pm 0.62$ & $3.62 \pm 0.47$ & $\mathbf{4.32 \pm 0.36}$ \\
    Navigation Confidence      & $2.48 \pm 0.78$ & $3.45 \pm 0.64$ & $\mathbf{4.12 \pm 0.51}$ \\
    Overall Preference         & $2.86 \pm 0.71$ & $3.58 \pm 0.55$ & $\mathbf{4.25 \pm 0.39}$ \\
    \bottomrule
  \end{tabular}
  }
  \vspace{-10pt}
\end{table}

Subjective MOS results in Table~\ref{tab:mos_results} further show that EscFOA receives the highest scores in perceived ease, listening comfort, navigation confidence, and overall preference. These findings provide preliminary evidence that geometry-consistent auditory cues can reduce compensatory effort and improve immersive learning experience.

\section{Conclusion}
In this study, we presented EscFOA, a learning-oriented, geometry-aware spatial audio framework for immersive 360-degree educational environments. EscFOA transforms scene geometry into \emph{acoustic scaffolding} and generates FOA aligned with environmental structure. Results suggest that EscFOA can better support spatial orientation and reduce perceived cognitive load than conventional audio representations under visually deprived conditions. Because it operates directly on 360-degree videos without specialized acoustic capture or manual annotation, EscFOA offers a practical pathway toward more accessible immersive learning. A limitation of this study is that the evaluation was conducted with blindfolded participants. Future work will involve visually impaired learners in broader educational settings and further explore universal design alignment.

\bibliographystyle{IEEEtran}
\bibliography{references} 

\begin{thebibliography}{10}
\providecommand{\url}[1]{#1}
\csname url@samestyle\endcsname
\providecommand{\newblock}{\relax}
\providecommand{\bibinfo}[2]{#2}
\providecommand{\BIBentrySTDinterwordspacing}{\spaceskip=0pt\relax}
\providecommand{\BIBentryALTinterwordstretchfactor}{4}
\providecommand{\BIBentryALTinterwordspacing}{\spaceskip=\fontdimen2\font plus
\BIBentryALTinterwordstretchfactor\fontdimen3\font minus \fontdimen4\font\relax}
\providecommand{\BIBforeignlanguage}[2]{{%
\expandafter\ifx\csname l@#1\endcsname\relax
\typeout{** WARNING: IEEEtran.bst: No hyphenation pattern has been}%
\typeout{** loaded for the language `#1'. Using the pattern for}%
\typeout{** the default language instead.}%
\else
\language=\csname l@#1\endcsname
\fi
#2}}
\providecommand{\BIBdecl}{\relax}
\BIBdecl

\bibitem{who2023blindness}
\BIBentryALTinterwordspacing
{World Health Organization}, ``Blindness and vision impairment,'' WHO Fact Sheet, Aug. 2023. [Online]. Available: \url{https://www.who.int/news-room/fact-sheets/detail/blindness-and-visual-impairment}
\BIBentrySTDinterwordspacing

\bibitem{chen2025vf}
X.~Chen, D.~Han, Q.~Qu, and Y.~Shen, ``Vf-lens: Enhancing visual perception of visually impaired users in vr via adversarial learning with visual field attention,'' in \emph{2025 IEEE Conference Virtual Reality and 3D User Interfaces (VR)}.\hskip 1em plus 0.5em minus 0.4em\relax IEEE, 2025, pp. 420--430.

\bibitem{lahav2012virtual}
O.~Lahav, D.~Schloerb, S.~Kumar, and M.~Srinivasan, ``A virtual environment for people who are blind--a usability study,'' \emph{Journal of assistive technologies}, vol.~6, no.~1, pp. 38--52, 2012.

\bibitem{xu2023wearable}
P.~Xu, G.~A. Kennedy, F.-Y. Zhao, W.-J. Zhang, and R.~Van~Schyndel, ``Wearable obstacle avoidance electronic travel aids for blind and visually impaired individuals: A systematic review,'' \emph{IEEE Access}, vol.~11, pp. 66\,587--66\,613, 2023.

\bibitem{zotter2019ambisonics}
F.~Zotter and M.~Frank, \emph{Ambisonics: A practical 3D audio theory for recording, studio production, sound reinforcement, and virtual reality}.\hskip 1em plus 0.5em minus 0.4em\relax Springer, 2019.

\bibitem{picinali2014exploration}
L.~Picinali, A.~Afonso, M.~Denis, and B.~F. Katz, ``Exploration of architectural spaces by blind people using auditory virtual reality for the construction of spatial knowledge,'' \emph{International Journal of Human-Computer Studies}, vol.~72, no.~4, pp. 393--407, 2014.

\bibitem{wood1976role}
D.~Wood, J.~S. Bruner, and G.~Ross, ``The role of tutoring in problem solving,'' \emph{Journal of child psychology and psychiatry}, vol.~17, no.~2, pp. 89--100, 1976.

\bibitem{luo2026dynfoa}
Z.~Luo, L.~Chen, Q.~Qu, X.~Chen, and Y.~Shen, ``{DynFOA}: Generating first-order ambisonics with conditional diffusion for dynamic and acoustically complex 360-degree videos,'' \emph{arXiv preprint arXiv:2602.06846}, 2026.

\bibitem{kerbl20233d}
B.~Kerbl, G.~Kopanas, T.~Leimk{\"u}hler, and G.~Drettakis, ``3d gaussian splatting for real-time radiance field rendering.'' \emph{ACM Trans. Graph.}, vol.~42, no.~4, pp. 139--1, 2023.

\bibitem{Liu2023AudioLDM}
H.~Liu, Z.~Chen, Y.~Yuan, X.~Mei, X.~Liu, D.~Mandic, W.~Wang, and M.~D. Plumbley, ``{AudioLDM}: text-to-audio generation with latent diffusion models,'' in \emph{Proceedings of the 40th International Conference on Machine Learning}, 2023, pp. 21\,450--21\,474.

\bibitem{wilson2000situated}
B.~G. Wilson and K.~M. Myers, ``Situated cognition in theoretical and practical context,'' \emph{Theoretical foundations of learning environments}, pp. 57--88, 2000.

\bibitem{zou2023spatial}
X.~Zou and Y.~Zhou, ``Spatial cognition of the visually impaired: A case study in a familiar environment,'' \emph{International Journal of Environmental Research and Public Health}, vol.~20, no.~3, p. 1753, 2023.

\bibitem{liu2025omniaudio}
H.~Liu, T.~Luo, K.~Luo, Q.~Jiang, P.~Sun, J.~Wang, R.~Huang, Q.~Chen, W.~Wang, X.~Li \emph{et~al.}, ``{OmniAudio}: Generating spatial audio from 360-degree video,'' in \emph{International Conference on Machine Learning}.\hskip 1em plus 0.5em minus 0.4em\relax PMLR, 2025, pp. 39\,060--39\,084.

\end{thebibliography}

\end{document}